# New Symbols for Base-16 and Base-256 Numerals

MacKenzie Cumings
Grand Rapids, Michigan, USA
mackenzie.cumings@gmail.com

Valdis Vītoliņš
Vidzeme University of Applied Sciences,
Valmiera, Latvia
valdis.vitolins@odo.lv

*Abstract* — **A new system of hexadecimal and base-256 numerals is proposed whose digit shapes are based on binary numerals. The proposed numerals are implemented in open source fonts and integrated into popular editors (Notepad++ and Eclipse) to prove the concept.**

## I. Introduction

Hexadecimal numerals conventionally are permutations of 16 symbols, where the symbols are the digits '0'-'9' and the letters 'A'-'F'. This representation is widely used but not universally accepted, at least in that there are some who have proposed other sets of symbols as an improvement on the conventional symbols. This paper is one such proposal; it proposes a new set of 16 symbols that have binary numerals embedded in their shape. Each pair of these symbols can be combined into one symbol as a typographic ligature, the result being a set of 256 symbols that is suitable for identifying bytes in computer memory and also for representing numbers as base-256 numerals.

In addition to presenting a new set of symbols, this paper attempts to demonstrate that these symbols are designed well and are useful in practice. To that end, it identifies other proposed sets of symbols for hexadecimal numerals, defines criteria that are desirable for a set of hexadecimal symbols, and compares the various proposals against the criteria. It attempts to justify some of the decisions made in designing the proposed set of symbols. Finally, it points to two existing software applications that use the proposed symbols to depict the binary content of computer files.

Notes on terminology: In this paper, a "number" is an arithmetical value. A "numeral" or "positional numeral" is a sequence of symbols that denote a number. A "digit" is a symbol that is a component of a numeral. To help prevent confusion of numbers themselves with numerals and symbols that represent them, numbers shall be referred to by their English names (e.g. ten, eleven, twelve) or by numerals (e.g. 10, 11, 12); symbols shall be referred to by surrounding them with single quotes (e.g. '0', 'F'), and numerals shall be referred to by surrounding them with double quotes (e.g. "100"). So 8 is the number eight, '8' is a symbol for the number eight that may be part of a numeral and "18" is a numeral for the number eighteen.

In this paper, positional numerals composed of the digits '0'-'9' and the letters 'A'-'F' shall be referred to as Standard Hexadecimal. Sets of 16 symbols that consist of the digits '0'-'9' and 6 additional symbols shall be referred to as Supplemented Digits. Sets of symbols that depict binary representations of the numbers that they symbolize shall be referred to as Binary-Encoding Symbols and the quality of depicting a binary numeral in one way or another shall be referred to as Binary Encoding.

Base-256 numerals shall simply be called "base-256". We find no word for such numerals. If such a word is ever desired, we recommend "sedecisedecimal", which is derived from the Latin word sedecim and means "sixteen sixteens".

Standard Hexadecimal numerals are prefixed with "0x" to distinguish them from decimal numerals. This is a common convention in computing.

## II. Prior Art

Throughout the history of civilization, many different numeral systems have been used, but gradually the positional numeral system[1] of Arabic[2] numerals has come to dominate. The shape of its digits is arbitrary, but there are numeral systems where the shape of digits depict the numbers they signify (with exceptions like zero), such as the Mayan[3], Babylonian[4], Korean Kanja and Hangul[5] systems, or mostly depict the numbers they signify, such as Braille[6].

---

[1] https://en.wikipedia.org/wiki/Positional_notation
[2] https://en.wikipedia.org/wiki/Arabic_numerals

The earliest proposal for hexadecimal numerals that we find was published in 1862. We find no other proposals until the early days of computers. In that era, more than one set of hexadecimal symbols were introduced. In addition to Standard Hexadecimal, which was introduced by IBM in 1963, the Bendix G-15 computer used '0'-'9' and 'U'-'Z'. There are others. All sets of symbols we find from this era are the decimal digits supplemented with additional characters.

Since Standard Hexadecimal became the de facto standard, a few other sets of hexadecimal symbols have been proposed, sometimes in response to perceived faults of Standard Hexadecimal. Most of the recent proposals we have found are informal postings on the Internet, and we find no academic papers published on the subject. Thus it appears that creating new system of hexadecimal numerals has not seriously been considered by computer scientists. Even so, the proposals from enthusiasts are evidence that not everyone is happy with Standard Hexadecimal and thus improvements may be interesting to some.

The following is a list of proposals for hexadecimal symbols, in chronological order:

- 1862. John Nystrom proposes a "Tonal system" as a more practical alternative to the decimal system. He assigns the usual Arabic digits to numbers 0 through 8, assigns the digit '9' to ten, and invents new symbols for numbers 9, 11, 12, 13, 14 and 15 [1].

- 1968. Bruce Martin proposes a system of Binary-Encoding Symbols in a letter to the editor of Communications of the ACM [2].

- 1969. Robert Lapointe registers the invention of a "Bibi-binary" system for hexadecimal numbers as different shapes that connect four dots [3].

- 2009: A blogger named "mejd" proposes replacements for the Standard Hexadecimals' 'A'-'F' in his blog. The replacements are reminiscent of both 'A'-'F' and the octal numerals for the numbers that they represent.

- 2009: Users on a website forum devoted to dozenal (Base-12) numerals discuss hexadecimal numerals, suggesting that Standard Hexadecimal could be altered to use the same symbols as dozenal numerals.[7]

- 2009: In a blog post, one of the authors of this paper proposes Binary-Encoding Symbols that can be depicted on a seven-segment display[8].

- 2011: In a blog post, Woody Thrower proposes a set of Binary-Encoding Symbols called "Hexy Digits"[9].

- 2012: A user named "Trismarck" comments on a blog post, proposing several sets of Binary-Encoding Symbols that also can be shown on a seven-segment display[10].

- 2012: In a blog post, Andrea Faulds takes issue with 'A' representing an even number when the ASCII code for 'A' is odd, 'B' representing an odd number when the ASCII code for 'B' is even, etc., and proposes the symbols '0123456789XABCDE' instead[11].

- 2015: In a blog post, one of the authors of this paper proposes Binary-Encoding Symbols that can be depicted on a seven-segment display and also can be combined in ligatures to form symbols for base-256 numerals[12].

III. DESIRABLE QUALITIES FOR A SET OF SYMBOLS

If anyone proposes a new set of hexadecimal symbols, it is only reasonable to ask if it is any good, and why use this set and not another? To answer these questions, we have defined several qualities that we believe are desirable for a set of hexadecimal symbols to have. Any proposed set of symbols may be evaluated for these qualities (the coded names for these qualities are used in *Table 1*).

---

[3] https://en.wikipedia.org/wiki/Maya_numerals

[4] https://en.wikipedia.org/wiki/Babylonian_numerals

[5] https://en.wikipedia.org/wiki/Korean_numerals

[6] https://en.wikipedia.org/wiki/Braille

[7] http://z13.invisionfree.com/DozensOnline/ar/t331.htm

[8] http://mackwai.blogspot.com/2009/07/alternative-hexadecimal-digits.html

[9] http://woodpress.org/2011/10/17/hexy-digits/

[10] https://www.blogger.com/comment.g?blogID=35719397&postID=3545068850225841741&bpli=1&pli=1

[11] http://blog.ajf.me/2012-03-31-hexadecimal

[12] http://odo.lv/Blog/150502?language=en

- **MNE**: *The symbols should be mnemonic*. There should be some aspect of the symbols that assists readers and writers in associating them with the numbers that they represent.

- **STR**: *Each symbol should require no more than 2 pen strokes to write*. The digits should be easily writable by hand.

- **LIG**: *The symbols should be combinable within ligatures to be symbols for base-256 digits*. Hexadecimal symbols are commonly used in computing, often in contexts where the values of bytes of memory are being inspected. In spite of this, each value of byte does not have its own symbol. Byte values are typically represented as pairs of hexadecimal symbols. Why not combine these pairs and create a distinct symbol for each value of byte?

- **AMB**: *There is small chance of ambiguity due to quickly-drawn digits*. No two symbols or ligatures should resemble each other closely as it is important that they be distinguishable. Note that this quality may conflict with *MNE*, because numbers with similar values should look similar if strictly positional rules are used.

- **DSP**: *The symbols can be depicted with a seven-segment display*. The seven-segment display is the conventional display for digital clocks and other devices with little computing resources. It is a very common electronic component, so it is a definite advantage for a set of symbols to be compatible with it.

- **BIN**: *The symbols are Binary-Encoded*. In computing, inspecting the values of bits in memory and inspecting bytes often go together, so it is helpful for the binary version of a byte, and therefore its bits, to be visible. It also provides those who know binary numerals with a means of determining the values of the symbols without relying on rote memorization.

- **0**: *The shape of the 0 digit is preserved*. Binary, decimal, octal and Standard Hexadecimal numerals all use the same symbol for 0.

- **1**: *The shape of the 1 digit is preserved*. Binary, decimal, octal and Standard Hexadecimal numerals all use the same symbol for 1.

- **TRN**: *The digits can be translated to and from binary easily*. Different sets of Binary-Encoding Symbols may be more or less easy to translate to and from binary. The easier it is to translate, the more helpful the encoding will be in enabling readers or writers to associate bits with bytes and symbols with numbers.

## IV. Evolution of Ideas for Hexadecimal Digits

To encode the numbers 0 through 15 as binary, 4 bits are needed. The simplest form for symbols that encode 4 bits is found as follows: The symbols have to show the presence or absence of each bit in each symbol, with the exception of the symbol for zero. Zero is excepted because, if zero is represented as an empty space, the numerals would not be easily readable. Although each bit could be represented by a dot, such as in Braille, writing proper dots takes more effort than writing strokes. Thus, the simplest form for easily-writable symbols that encode 4 bits would be four "Boolean sticks" in the shape of a rectangle. However, such an arrangement would still not be easily readable, so Robert Laponte proposed a more effective format [3]:

Figure 1. hexadecimal digits proposed by Robert Laponte

These digits are easily recognizable and the shapes are systematic, but it is hard to see how they encode bits. The digits proposed by Bruce Martin are exactly positional, where the bits are encoded as ordered horizontal lines, joined by vertical line for recognizability [2]:

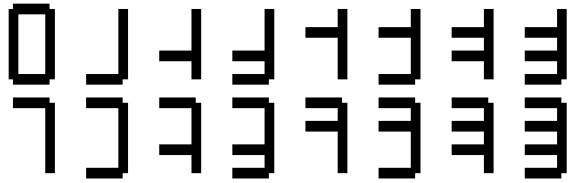

Figure 1.  Hexadecimal digits proposed by Bruce Martin

Unfortunately, these numerals lack ease of writing, because they have to be written with up to five strokes, while decimal numerals can be written with (more or less sophisticated) two strokes. Also, one of the middle strokes may be confused with the other if it is written too close to the middle.

The digits we are proposing are in *Figure 3*. They are based on these premises:

- The binary encoding of each digit should be clearly visible.
- Each hexadecimal digit should be writable by hand with no more than two strokes.
- If right- and left-sided vertical "sticks" are introduced to depict one of the bits, we can use fewer horizontal strokes, which makes the shapes of numerals easier to write and less likely to be ambiguous.
- The number a symbol encodes is the sum of these powers of two.

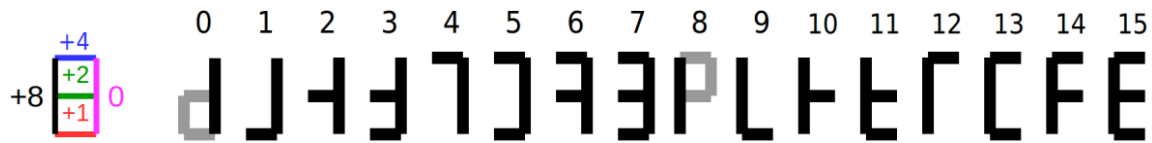

Figure 2.  Bit layout of the proposed hexadecimal digits

The vertical lines (+0 and +8 bits) are exclusive and work as triggers. The digits for 0 and 8 do not follow the bit layout exactly, since both could not be represented as vertical lines without creating ambiguity, and representing one or the other with a vertical line might cause it to be confused with the number 1, therefore their shape is complemented with more lines (show in grey in *Figure 3*). With this bit layout, no digit has:

1. Less than one line altogether,
2. Two vertical lines (besides exceptions for zero and eight),
3. More than four lines in altogether.

For these binary-encoding symbols, "sticks arithmetic" can be used to add the values of binary digits with an obvious method carrying binary digits. This may aid in teaching the values of the symbols.

⌐ + ⌐ = ¬
⌐ + ¬ = ╕
⌐ + ┐ = ╗
⌐ + ⌊ = ├
⌐ + ⌈ = ⌞

Figure 3.  Bit arithmetic of the proposed digits

To improve readability, zero and eight may be represented in one of following forms:

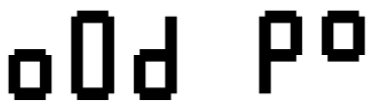

Figure 4.  Possible shapes of zero and eight for the proposed numerals

V. Development of Base-256 Digits using Ligatures of the Proposed Hexadecimal Digits

While designing the symbols, we found that the exceptional shapes for zero's digit and eight's digit improved the readability of the ligatures that are produced by joining two digits together. We decided to show zero as "bottom zero" and eight as "top zero". *Figure 6* shows hexadecimal digits in range from 0x0 to 0xF and a table of the base-256 digits in their full range of 0x00 to 0xFF.

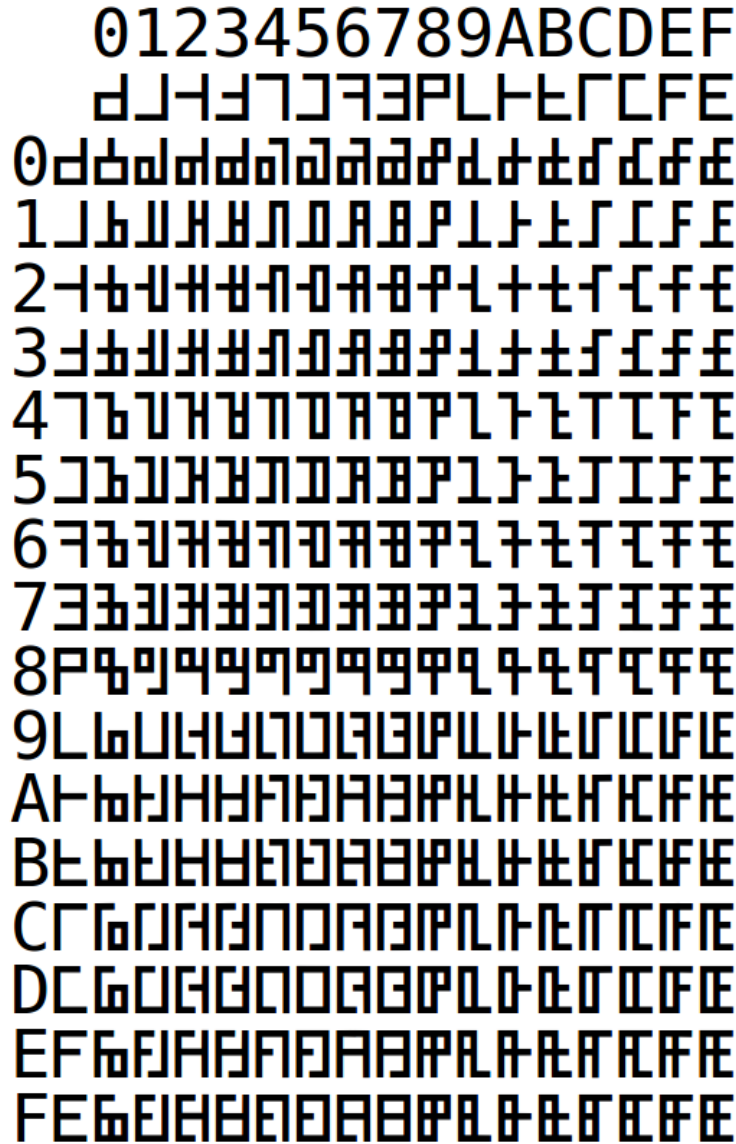

Figure 5. *Proposed hexadecimal digits and their ligatures for base-256 digits*

Note that the proposed base-256 digits are shown with the same width and margin as standard typographic characters. Also note that they can be shown in a 12-cell digital matrix without losing the visibility of the encoded bits.

VI. Readability and Writeability of the Proposed Digits

A. *Writeability*

The handwritten versions of these digits may depart from the stick matrix in order to reduce the number of strokes needed to write them. This can be done without losing the visibility of the bit-encoding, as can be seen in *Figure 7*. The ligatures for base-256 numerals can be handwritten using similar principles.

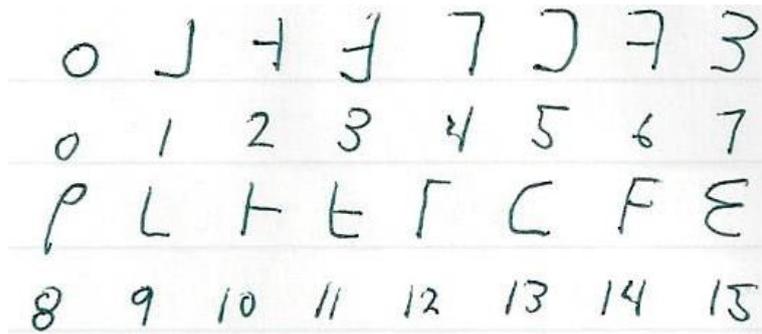

Figure 6. *Handwritten example of the proposed digits*

*B. Readability*

As humans are best at seeing patterns and deviations from common patterns, a purely positional representation of numerals could be too monotonous and could weaken the recognition of digits in long numbers. To make the digits more distinguishable, different shapes for the digits were developed. After analyzing decimal digits, the following features were considered for making the shapes of numerals and ligatures more distinct:

1. Sharp edges/rounded edges of corners,
2. Straight/curved lines,
3. Rectangular/sloped lines,
4. Prolonged shapes (slight subscript/superscript) of numerals.

For example, the proposed digit for 15 can be written with the following shapes: 'E', 'Ɛ' or '€'. Each of them has the same "topology" (i.e. placement of lines that depict bits), but each has a different graphical shape. In our proposal, rules for shape deviation is also applied positionally in a systematic manner.

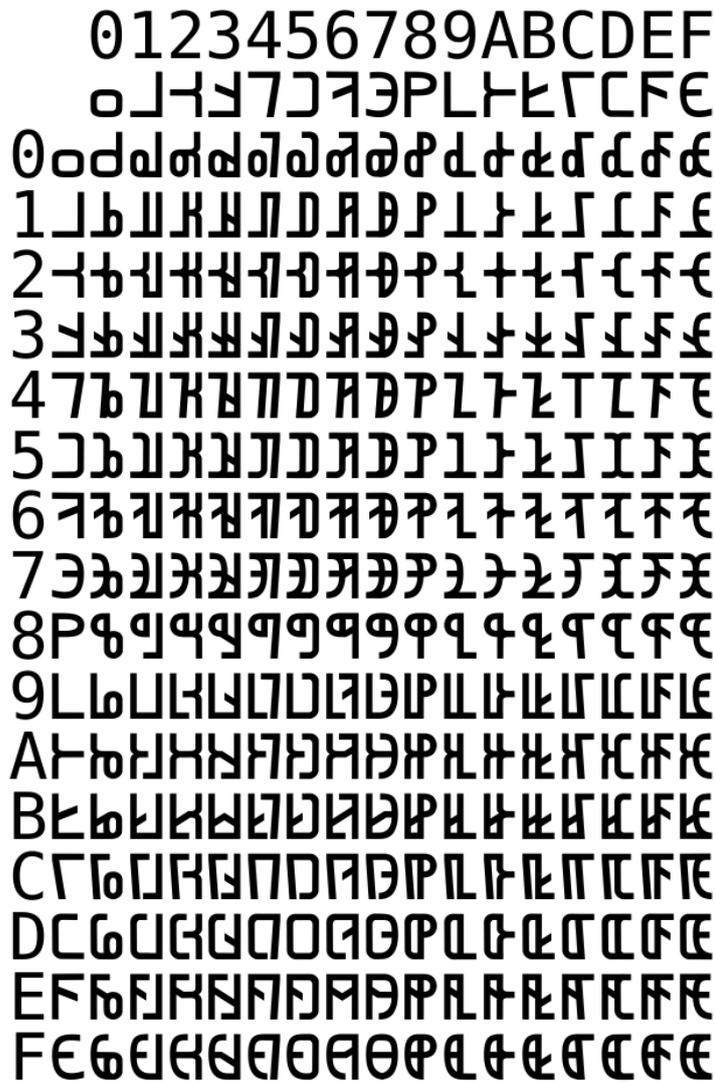

Figure 7. *Examples of shape deviations for various digits in the proposed base-256 digit set*

VII. COMPARISON OF NUMERAL SYSTEMS

*Table 1* shows a comparison of the various proposals mentioned in *Section II* according to the qualities defined in *Section III*. In order to keep the comparison brief, we include only two of the Supplemented Digit sets in the comparison — Standard Hexadecimal and MEJD's symbols — because we find none of the other sets to be noticeably better than both of these sets.

TABLE I. COMPARISON OF QUALITIES OF PROPOSED NUMERALS

| Symbol Set | MNE | STR | LIG | AMB | DSP | BIN | 0 | 1 | TRN | Score |
|---|---|---|---|---|---|---|---|---|---|---|
| Standard Hexadecimal '0'-'9' 'A'-'F' |  | ✓ |  | ✓ | ✓ |  | ✓ | ✓ |  | 5 |
| Martin 1968 | ✓ |  |  |  |  | ✓ |  |  | ✓ | 3 |
| Laponte 1969 | ✓ | ✓ |  | ✓ |  | ✓ | ✓ | ✓ |  | 6 |
| MEJD 2009 | ✓ |  |  | ✓ |  |  | ✓ | ✓ |  | 4 |
| Cumings 2009 | ✓ | ✓ |  | ✓ | ✓ | ✓ | ✓ | ✓ |  | 7 |
| Hexy Digits 2011 | ✓ |  |  | ✓ |  | ✓ | ✓ | ✓ |  | 5 |
| Trismarck 2012 |  |  |  |  | ✓ | ✓ |  |  |  | 2 |
| Vītoliņš 2015 | ✓ | ✓ | ✓ |  | ✓ | ✓ | ✓ |  | ✓ | 7 |

| Vītoliņš and Cumings 2017 | ✓ | ✓ | ✓ | ✓ | ✓ | ✓ | ✓ |  | ✓ | 8 |

Of course, some qualities may be more important than others, and some sets may have qualities in differing degrees. For instance, *Martin 1968* is best for translation to and from binary, because of its strong positional principles. *Laponte 1969* is good for handwriting and unambiguous strokes, but its numerals poorly present bits. No digits except those *Vītoliņš 2015* can be used in joined ligatures to create base-256 digits, either because the digits are "too dense" for larger values (e.g. *Martin 1968*) or they become indistinct because of their shape.

VIII. IMPLEMENTATION OF PROPOSED NUMERALS

Hexadecimal numerals are often used in computing to represent the contents of computer memory. In fact, one tool used frequently by computer programmers is a *hex editor:* a software application that enables its user to view the full contents of a file, byte-by-byte, as a series of hexadecimal symbols, and make precise changes to the file's contents by typing hexadecimal symbols. The digits proposed in this paper have already been implemented in custom-built fonts and integrated into two hex editors:

1. As a plugin for Notepad++[13], which is a popular text editor
2. As a plugin for Eclipse[14], which is a popular IDE

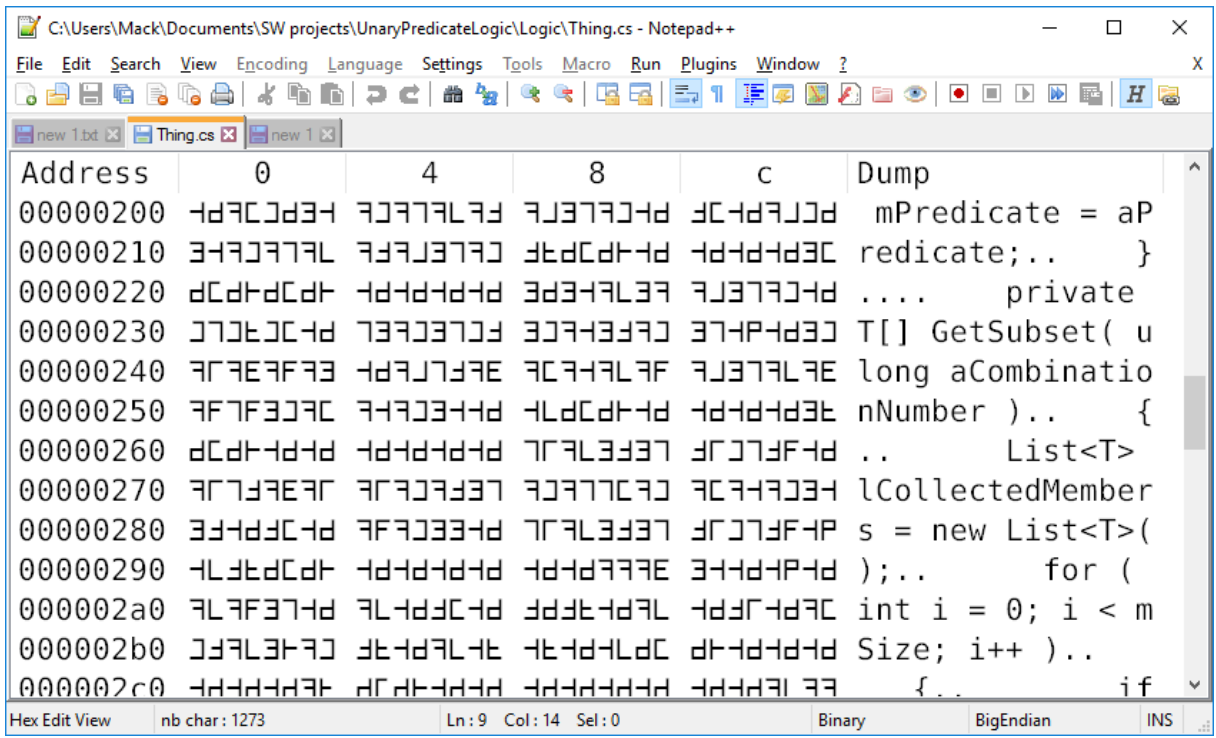

Figure 8.  *Screenshot of an "Alternative Hexadecimal Editor" with proposed numerals in Notepad++*

---

[13] http://somerby.net/mack/HexEditor.zip
[14] https://github.com/valdisvi/HexEditorPlugin/

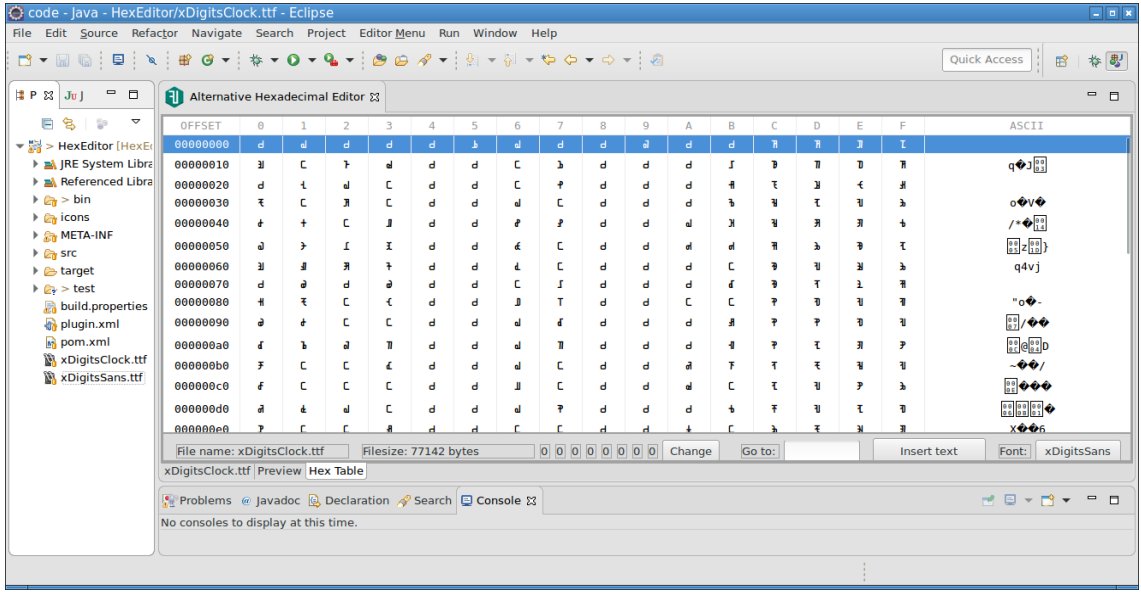

Figure 9. *Screenshot of an "Alternative Hexadecimal Editor" with proposed numerals in Eclipse*

## IX. CONCLUSION AND FUTURE WORK

Our proposed hexadecimal numerals provide a good balance of binary encoding, simplicity, redundancy, readability and writeability. The binary-encoding scheme, with the reasonable exceptions for 0 and 8, allows to the digits to be joined together in "ligatures" and represent the value of an 8-bit byte with a single digit.

Future work is necessary in the area of usability. It should be investigated what the best approach is to make the exact shape of the symbols not only distinct but also ergonomic. This system of numerals should be popularized in computer science studies and applied fields.

Names of proposed numerals could also be binary-encoding, such as what was proposed by Robert Laponte [3]: Ho, Ha, He, Hi, Bo, Ba, Be, Bi, Ko, Ka, Ke, Ki, Do, Da, De, Di. For base-256 digits names are joined in two syllable words such as: Hoha, Hehi, Boba, Bebi, Koka, Keki, Doda, Dedi. It should be investigated which consonant-vowel pairs are the most international and allows adding suffixes for names for languages with inflexions.

REFERENCES

[1] J. W. Nystrom, "Project for a New System of Arithmetic, Weight, Measure and Coins Proposed to be Called the Tonal System with Sixteen to the Base", J. B. Lippincott & Co, Philadelphia, Trubner & Co, London, 1862.

[2] B. A. Martin, "Letters to the editor: On binary notation", Communications of the ACM, Volume 11 Issue 10, Oct. 1968, Page 658.

[3] Brevet d'invention n° 1.569.028, Procédé de codification de l'information, Robert Jean Lapointe, demandé le 28 mars 1968, délivré le 21 avril 1969.